\documentclass[a4paper]{article}
\usepackage[ansinew]{inputenc}
\usepackage[T1]{fontenc}
\usepackage[UKenglish]{babel}
\usepackage{amsmath}
\usepackage{graphicx}
\usepackage{amssymb}
\usepackage{siunitx}
\usepackage{cite}
\usepackage{epstopdf}
\usepackage{subcaption}
\usepackage{color}
\usepackage{cleveref}
\usepackage{geometry}
\usepackage[singlespacing]{setspace} 
\usepackage{longtable}
\usepackage{bm}
\usepackage[table]{xcolor}

\begin{document}

\bibliographystyle{unsrt}
\title{Remote Analysis of Femoroacetabular Impingement by a triade of label-free optical spectroscopy techniques}
\author{Martin Hohmann$^{1,2,*,***}$,Lucas Kreiss $^{2,3,**,***}$, Faramarz Dehghani$^{4}$,\\ Dongqin Ni$^{1,2}$, Max Gmelch$^{2}$, Oliver Friedrich $^{2,3}$,
	Lorenz Büchler$^{5}$,\\ Michael Schmidt$^{1,2}$}
\maketitle
\noindent
$^1$ Institute of Photonic Technologies (LPT), Friedrich-Alexander-Universität Erlangen-Nürnberg (FAU), Konrad-Zuse-Straße 3/5, 91052 Erlangen, Germany\\
$^2$ Erlangen Graduate School in Advanced Optical Technologies (SAOT), Paul-Gordon-Straße~6, 91052 Erlangen, Germany \\
$^3$ Institute of Medical Biotechnology (MBT), Friedrich-Alexander-Universität Erlangen-Nürnberg (FAU), Paul-Gordan-Straße 3, 91052 Erlangen, Germany\\
$^4$ Institut für Anatomie und Zellbiologie, Martin Luther Universität Halle-Wittenberg, Große Steinstraße 52, 06097 Halle (Saale), Germany \\
$^5$ Department for Orthopaedics and Traumatology, Tellstrasse 25, 5001 Aarau, Switzerland\\
*Martin.Hohmann@FAU.de,  **Lucas.Kreiss@fau.de\\
*** First Author

\section*{Abstract}

This paper introduces the combination of Laser-induced breakdown spectroscopy (LIBS), Raman spectroscopy (RS) and diffuse reflectance spectroscopy (DRS) in the field of biomedical research. Thereby, the results from RS and LIBS are combined with previous DRS results. These non-invasive optical methods, used together, offer thorough analysis of the absorptive and scattering behaviour, elemental composition, and molecular bindings of tissue, without resulting in considerable harm or changes to the sample or the requirement of markers. The study centres on applying this triad of techniques to tissues affected by Femoroacetabular Impingement (FAI), a condition characterised by reduced hip mobility due to developmental deformities of the hip joint. The research results in a biochemical pathway model of the condition causing the red staining caused by FAI which origin was unknown until now. This proves that this approach may have significant implications for numerous medical applications and support in exploring complex chemical and biochemical pathways.

\section{Introduction}

For numerous biomedical problems, it is extremely advantageous to be able to merge information from different underlying aspects. Hence, a combination of technologies is needed to achieve this. Ideally, all technologies ought to be easy-to-use and have the potential for \textit{in vivo} applications. Furthermore, in many cases, a high spatial resolution is beneficial as well. Laser-induced breakdown spectroscopy (LIBS) alongside Raman spectroscopy (RS) and diffuse reflectance spectroscopy (DRS), as a complementary method, could offer a potential solution for measuring many different aspects of tissue under investigation. All techniques possess the benefits of remote optical technologies: they do not cause significant damage to the object and do not require a high level of protection of the tissue under investigation or patient and the examiner. In addition, the proposed combination does not require the use of biochemical labels or fixation techniques.

Here, we show the combination of LIBS, RS, DRS for the first time on human samples. That enables the collection of elemental and biochemical information in combination with information of the absorbers such as hemoglobin or melanin. The paper is structured as follows: First, information about LIBS is provided. Afterwards, RS is discussed. As we published the results for DRS on FAI already~\cite{kreiss2019diffuse}, only the findings that are relevant for FAI are briefly summarized and discussed in the according sections.

For a LIBS measurement, an initial high-power laser pulse ablates a microscopic portion of the sample and a plasma is generated from the material under investigation. The recombination of elements within the plasma then generates emission of light that can be collected and analyszed by a spectromter to measure the atomic and molecular~\cite{LIBS_molecule} emission lines. The research on LIBS, especially its ability to measure the elemental composition is drawing more and more attention during the last decades~\cite{LIBS_review}. LIBS allows the measurement of elemental concentrations in solids, liquids and gases~\cite{LIBS_review}. Even absolute quantification is possible with a precision of a few parts per million~\cite{LIBS_precision}.  In the last years, LIBS is becoming more commonly used for medical applications with extremely promising results~\cite{skalny2023medical}.

RS targets the vibrational and rotational energy bonds of molecules in the sample. It is a widely used method in many fields~\cite{RS_review,orlando2021comprehensive}, such as the assessment of food quality~\cite{RS_review_food_quality}, the analysis of carotenoids in biological matrices~\cite{RS_review}, or in cancer research~\cite{RS_review2,RS_EpithelialCancer,RS_BreastCancer}. Normally, the Raman-spectra of organic molecules are analysed by specific peaks~\cite{orlando2021comprehensive}. For example, it could be shown that the concentration amid I of significantly increases with age in patients compared to amid III in human bone~\cite{RS_bone_aging}. In this study, the combination of LIBS and RS is applied to tissue effected by Femoroacetabular Impingement~(FAI)~\cite{FAI}.

FAI defines a reduced range of motion of the hip and early bony contact of the femur with the acetabulum during movement. In many cases, the cause of the reduced range of motion is a developmental deformity of the hip joint with an increased size of the femoral neck, a deep acetabulum, or a combination of both. It is also likely that symptomatic FAI is a risk factor for the later development of hip osteoarthritis~\cite{FAI,FAI_arthrose,FAI_arthrose2,FAI_arthrose3,FAI_arthrose4,FAI_arthrose5}. State of the art treatment for FAI is to restore normal hip anatomy to regain normal range of motion of the hip, to eliminate symptoms and prevent future development of osteoarthritis. In open surgery, the entire cartilage of the femoral head often appears normal immediately after dislocation. In open surgery, the entire cartilage of the femoral head often becomes reddened when exposed to air, upon dislocation~\cite{kreiss2019diffuse}. In contrast to that, healthy cartilage, unaffected by FAI, remains whitish upon air-contact. This region is of paramount clinical interest as it defines the region of osteophyte development in advanced osteoarthritis. However, the origin of the reddening is not known.

\begin{figure}[ht!]
	\centering
	\includegraphics[width= \linewidth]{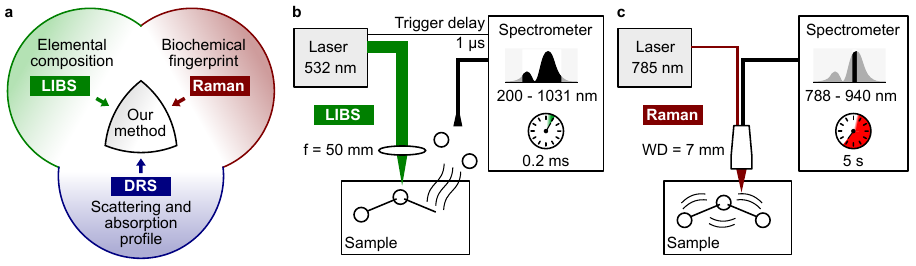}
	\caption{Schematic of the measurement set-ups.\\ a) Visualization of the combination of the different techniques. \\ b) RS set-up: The sample is positioned at the working distance of the probe, a diode laser is used for the excitation, and a spectrometer detects the back scattered light.\\ c) LIBS set-up: A frequency-doubled, Nd:YAG laser is focused at the surface of the sample and ablates a few micrometers of material. The recombination from the generated plasma cloud is detected as LIBS signal.}
	\label{setup}
\end{figure}

In this study, the capabilities in investigating FAI-affected tissue of three chosen technologies, namely RS and LIBS combined with DRS results from our previous study~\cite{kreiss2019diffuse} are discussed as shown in figure~\ref{setup}a. While the combination of LIBS and Raman is performed since long time~\cite{wiens2005joint_Raman_Libs}, their application especially in the medical field or for biochemical analysis is nearly not existant according to a review from 2021 from Dhanada et al.\cite{vs2021hybrid}. In this study, results LIBS and RS combined with previous DRS results, which is summarized in figure~\ref{setup}. This combination is used to investigate the differences between healthy and FAI-affected tissue. Based on these results, we propose a biochemical pathway that explains the appearance of red staining when FAI-affected tissue is exposed to air. It is possible to apply the same approach of combining LIBS, RS and DRS to many biomedical or biomedical problems in future.

\section{Results}

The results section consists of four parts. The first two present the outcomes for LIBS and Raman. The third part offers a discussion of the conclusions drawn from the first two sections combined with the pertinent findings from our prior DRS investigations. In the last part, a model for the molecular changes in tissue during FAI development is formulated.

\subsection{Laser-Induced Breakdown Spectroscopy (LIBS)}

\begin{figure}[ht!]
	\centering
	\includegraphics[width= \linewidth]{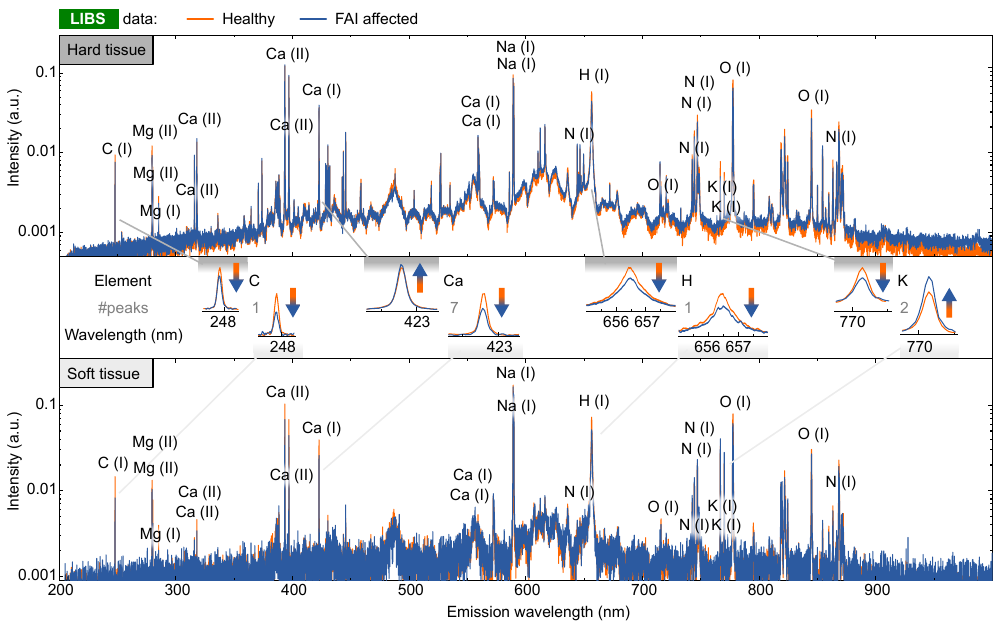}
	\caption{Averaged LIBS spectra of FAI affected and healthy reference tissue. All peaks used for further analysis are labelled. Top: The averaged spectrum from data sorted as hard tissue. Center: Example peaks for comparison. Bottom: The averaged spectrum from data sorted as soft tissue.}
	\label{LIBS_spectra}
\end{figure}

This section shows the mean spectra and the summarised results. The results of the analysis for each individual peak are given in the appendices (table~\ref{tab_LIBS_attachment_I} and \ref{tab_LIBS_attachment_II}). The mean spectra are shown in figure~\ref{LIBS_spectra}. The peaks used for data analysis are labelled. Generally, for LIBS, the majority of peaks arise from calcium owing to the large number of peaks. It is anticipated that all calcium peaks will exhibit the same pattern; however, only seven representative peaks are chosen for data analysis. 

Figure~\ref{LIBS_spectra} displays the contrast between healthy and FAI-affected tissue. The hard tissue of FAI-affected patients presents a higher concentration of calcium than healthy individuals. All calcium peaks exhibited this behaviour. This shows increased levels of hydroxyapatite, the Ca-based main constituent of hard tissue, in FAI-affected samples which is in accordance with MRI measurements~\cite{hydroxylapatite_FAI,hydroxylapatite_FAI2} demonstrating increased hydroxyapatite levels in FAI-affected tissue. Both findings support a higher amount of calcium.  When taking into account the reduction of carbon and hydrogen, it suggests a decrease in organic material in the FAI-affected tissue. The oxygen tends to be decreased in FAI-affected tissue despite its large presence in hydroxyapatite due to the reduced organic compounds in the FAI-affected tissue. Four out of five peaks demonstrate a decrease in nitrogen in the FAI-affected tissue, providing additional support for this conclusion. However, this element yields conflicting results and thus only provides minor additional support for the reduction of organic compounds. Potassium levels follow the same pattern as organic matter and, since it is largely found in cells, it can be regarded as a metric for organic matter as well.

The significance for soft tissue is lower owing to a smaller number of measurements. FAI-affected soft tissues contain less carbon, calcium, and hydrogen. A smaller amount of carbon indicates less organic tissue, while the decrease of calcium suggests that the mineralisation of the hard tissue collects the calcium from surrounding soft tissue as the calcium in hard tissue is used to grow bones. Hydrogen levels are significantly lower in FAI-affected tissue, which is consistent with the carbon peak and also indicates a reduction in organic tissue. All the peaks analyzed show a significant difference, while the calcium peaks exhibit low significance. However, all three peaks show the same trend. The other differences between FAI and healthy tissue for the remaining peaks will not be further discussed, as the differences are negligible.

\subsection{Raman Spectroscopy (RS)}

\begin{figure}[ht!]
	\centering
	\includegraphics[width= \linewidth]{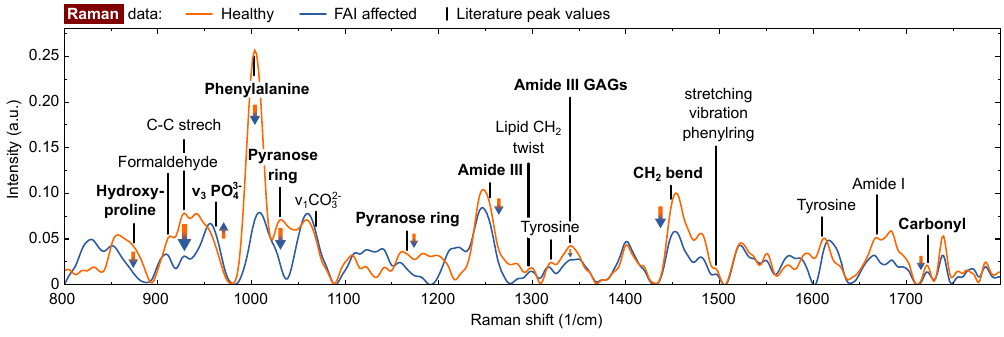}
	\caption{Mean Raman spectra of tissue affected by FAI and healthy reference tissue after denoising and baseline subtraction. All peaks for further analysis have been labelled, with the bold font indicating the peaks demonstrating significant differences (p<0.005 and influence of peaks is larger than of the sample number).}
	\label{Raman_spectra}
\end{figure}

% der 911 Raman peaks ist nicht bei 1 da die Peaks und nicht die Spektren normalisiert werden

Figure~\ref{Raman_spectra} depicts the averaged Raman spectra after denoising and baseline subtraction. Regardless of the statistical significance of each peak, there is an apparent decrease in almost all peaks that represent organic substances such as C-C stretch or pyranose ring in FAI affected tissues. Concurrently, the FAI affected tissue shows an increase in inorganic peaks ($v_3 PO_4 ^{3}$ and $\nu_1 CO_3^{2-}$). These findings align with those acquired from the LIBS data wherein there is a lower amount of organic matter present. Furthermore in literature, a decrease of cartilage cells is likely in FAI-affected tissue~\cite{olszewska2013activity}.

The increased levels of $v_3 PO_4 ^{3}$ and $\nu_1 CO_3^{2-}$ in FAI affected tissue are caused by the increased concentraion of hydroxyapatite~\cite{hydroxylapatite_Raman}. As an increase in the 1661\,\textrm{$cm^{-1}$} peak of amide I indicates collagen quality change caused by ageing~\cite{RS_bone_aging}, hydration/dehydration~\cite{RS_dehydration}, or radiological harm~\cite{RS_x_ray}, this observation demonstrates that the tissue affected by FAI is damaged. Nonetheless, it lacks statistical significance. Furthermore, there is a decline in the amount of hydroxyproline in FAI.

Additionally, the peak of phenylalanine is decreased in FAI-affected tissue. Phenylalanine is frequently transformed into tyrosine which, through 3,4-dihydroxyphenylalanine~(DOPA), produces melanin via enzymatic oxidation. Nevertheless, the peaks related to tyrosine do not indicate any substantial differences. Further, both pyranose ring peaks are significantly reduced in FAI affected tissue, much more than other organic substances. The same is true for phenylalanine. Consequently, the presence of these two peaks infers that further loss is attributable to other biochemical reactions.

\section{Diskussion and FAI-model}

This section is divided into two parts and primarily summarizes the findings from the LIBS and RS sections in conjunction with our prior DRS results~\cite{kreiss2019diffuse}. The first part examines the alteration of bone tissue due to FAI. By comparing with several other studies, it is evidenced that a comprehensive understanding can be achieved through the combination of DRS, Raman, and LIBS. The second section summarises the factors not accounted for by the bone model and combines them with established biochemical pathways to explicate the red discoloration of FAI-affected tissue upon exposure to air.

\begin{figure}[ht!]
	\centering
	\includegraphics[width= \linewidth]{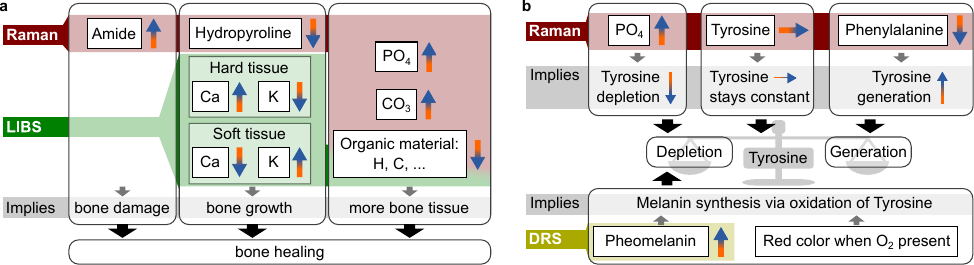}
	\caption{a) Bone growth model: Most of the findings are related to damage, grow and repair of the bone. b) FAI model: Biochemical model for the reason of the red coloring of the FAI-affected bone upon air contact.}
	\label{fig_FAI_model}
\end{figure}

According to figure~\ref{fig_FAI_model}a, alterations in FAI-affected tissue occur in three distinct categories. Firstly, elevated levels of amide I reveal bone damage, which is an outcome of frequent impingement. Secondly, reduced hydroxypyroline for collagen stabilization can be observed. The hydroxylation of proline by prolyl hydroxylase or other electron-withdrawing reactions greatly enhances collagen's conformational stability~\cite{szpak2011fish}. This indicates that the body responds to the damage by generating new bone. Moreover, there is an increase in calcium in hard tissue and a decrease in soft tissue, while the opposite holds true for potassium. Osteocytes predominantly store potassium intracellularly~\cite{chow1984electrophysiological}. As bones strengthen, they accumulate greater amounts of calcium from the surrounding soft tissue. Following fractures or microfractures, callus formation takes place, triggering a strong response from blood cells and osteoblasts. Consequently, this leads to the reinforcement or repair of the bone affected by FAI. As a third observation, there is an increase in the amount of ($PO_4 ^{3}$ and $CO_3^{2-}$) in the tissue affected by FAI, which also shows an increase of bone tissue. This may be attributed to the association between FAI and deposition of calcium hydroxyapatite at the osteochondral junction shown by MRI techniques~\cite{hydroxylapatite_FAI,hydroxylapatite_FAI2}. Additionally, LIBS indicates a lower concentration of organic tissue in general. Thus, there is more or hardened bone present. In summary, the bone damage, the body's attempt to repair it, and the resulting increased levels of inorganic tissue are evidently shown by our measurements.

The previous model elucidates the majority of the findings. However, the bone model alone cannot account for the red discoloration upon exposure to air, the strong reduction of phenyalanaline, and the presence of pheomelanin in FAI affected tissue which could be shown by our previous DRS study~\cite{kreiss2019diffuse}. To consider these points, the model has to be expanded, which is depicted in figure~\ref{fig_FAI_model}b. It is established that pheomelanin results from the enzymatic oxidation of Tyrosine catalysed by tyrosinase, through L-Dopa, Dopaquinone, and Cysteinyl DOPA~\cite{morris2002red,zaidi2014microbial}. First, tyrosinase catalyses the initial and rate-limiting step in the cascade of reactions that lead to melanin production from tyrosine~\cite{lai2017structure}. More precisely, tyrosinase hydroxylates tyrosine to DOPA and catalyses the oxidation of DOPA to DOPAquinone~\cite{lai2017structure}. Furthermore, the oxidation of Cysteinyl DOPA can occur under the appropriate conditions~\cite{ito1984oxidation}. Apart from this reaction chain, it is known that the growth of bone requires the presence of tyrosine. However, under typical conditions in the body, the concentration of tyrosine required for melanin synthesis is approximately 20 times higher than that necessary for bone growth~\cite{morris2002red}. As the measured amount of tyrosine remains constant, an increase in tyrosine levels is expected in the tissue affected by FAI. When there is sudden contact with air, the concentration of oxygen rises, which subsequently promotes the reaction of tyrosine to pheomelanin, and the excess of tyrosine is then used to produce pheomelanin. As a result, the tyrosine present in the bone affected by FAI reacts suddenly and produces pheomelanin, causing an alteration in the tissue's color to red.

\section{Materials and Methods}

\subsection{Patients}
The LIBS analysis was performed on 21 biopsies from 18 patients. One or two osteo-chondral samples from the anterior femoral head-neck junction were taken from each patient during open surgery for treatment of Cam Type FAI at the Orthopaedic Clinic of the University of Bern, Switzerland. The patients were given full information about the study. The patients then gave their informed consent to participate in the study. The study was conducted in accordance with the tenets of the Declaration of Helsinki and was approved by the Cantonal Ethics Commission of Bern (KEK - decision of 15 October 2015).

After the procedure, the samples were stored in a 4\% formaldehyde solution (Roti-Histofix~4\%, Carl Roth). Due to the coloring of the sample after contact with air, a clear identification of healthy and stained areas was easily possible. Therefore, histology was not performed. Each sample had the colored degeneration and an unaltered white border around it, which served as a healthy reference. From each of these samples, 5-10 measurement points were taken on the entire sample. It should also be noted that the LIBS measurements were acquired from bone as well as from cartilage tissue. An in-depth analysis of LIBS then allowed clear distinction between hard and soft tissue, based on molecular composition. This LIBS analysis was further supported by a Raman investigation of three additional samples.

Figure~\ref{setup}b shows the measurement set-ups used in this study. The LIBS measurement was performed on different samples than the Raman measurements. For all two modalities, the measurement points were selected randomly over the sample to minimise experimental bias.

\subsection{Laser-Induced Breakdown Spectroscopy (LIBS)}

A frequency-doubled Nd:YAG laser (Q-smart 450, Quantel Laser, Les ulis cedex, France) with a repetition rate of 10~Hz, a pulse duration of 5\,\textrm{ns} and a wavelength of 532\,\textrm{nm} was used for the LIBS measurements. The mean pulse energy measured 80\,\textrm{mJ}.  A lens with a focal length of 50\,\textrm{mm} was utilised to focus the beam just beneath the surface of the tissue sample. A 3D translation stage was also employed to relocate the measurement spot on the sample into the laser focus. To target the plasma cone produced by each laser pulse, the open tip of a UV-enhanced 50\,\textrm{$\mu m$} optical fibre was placed above the sample. The opposite end of the fibre was linked to a high-precision spectrograph (Mechelle ME 5000 Echelle, Andor, Belfast, UK). Equipped with an ICCD camera (A-DH334T-18F-03 USB iStar ICCD detector, Andor, Belfast, UK), the spectrometer has a spectral resolution ($\lambda / \Delta\lambda$) of 6000 within the 200~-~840\,\textrm{nm} spectral range. The laser was directly connected to the spectrograph to trigger the detector measurements with each laser pulse.  

All experiments were performed with a gate delay between the laser pulse and the detector of 1\,\textrm{$\mu s$} and a gate width of 0.2\,\textrm{ms}. In total, 21 samples from 18 patients were investigated with 6 to 10 spots per patient. 25 individual measurements were taken from each spot, with each measurement involving the ablation of a portion of the tissue. Thereby, first soft tissue and later hard tissue is measured by drilling into the tissue by laser ablation. Each ten LIBS pulses create a crater of approximately 300\,\textrm{$\mu m$} in depth and diameter~\cite{LIBS_fanuel}. Due to the irregular shape of the sample, it is possible that the ablation was not within the tissue, resulting in plasma generation from the air. To remove such measurements, the nitrogen content is compared with the $Ca^{2+}$ content. Nitrogen content is known to be very high in air, while $Ca^{2+}$ content is low. The ratio of the peak intensity of $Ca^{2+}$ at 392\,\textrm{nm} to N at 822\,\textrm{nm}~\cite{LIBS_NIST_ASD} was, therefore, used to filter out measurements in which the plasma had been generated in air. Spectra with a ratio exceeding 0.5 (n=412) were discarded for subsequent analysis. The identified peaks were ultimately assigned to elements using information derived from the NIST Atomic Spectra Database~\cite{LIBS_NIST_ASD}.

A comparable differentiation step was implemented in order to discern between hard and soft tissue areas. The tissue spectra vary significantly for both hard and soft tissue~\cite{LIBS_hard_soft_tissue}, hence a distinct analysis was executed for these two tissue categories. The distinctive structure of the molecular matrix has a significant impact on the ablation process, resulting in differences in plasma quantity, temperature and hence, the intensity of the recorded peaks. Therefore, a comparable separation must also be conducted for LIBS data. The classification was based on the Ca~(I) to K~(I) ratio recorded at 616\,\textrm{nm} and 766\,\textrm{nm}~\cite{LIBS_NIST_ASD}, which is known to exhibit the most significant variations between the cartilage and cortical bone~\cite{LIBS_hard_soft_tissue}. Spectra with a ratio below 0.5 are classified as soft tissue (cartilage), while spectra with a ratio above 1.0 are classified as hard tissue (cortical bone)~\cite{LIBS_hard_soft_tissue}. The spectra in between~(n=976) were ignored to ensure that they are correctly classified.

As a result, 965 spectra were obtained from the white reference tissue, and 1222 from the FAI affected tissue. Among these spectra, 885 and 1153 were hard tissue for the white reference and FAI affected tissue, respectively. The remaining spectra comprised of 80 soft tissue spectra from the white reference tissue and 69 from the FAI affected tissue. There are much less soft tissue spectra due to the fact that most laser pulses drilled deeper into the bone and the cartilage was relatively thin in most cases.

%The baseline was calculated by using a broad median filter with an order of 25 in combination with a smoothing algorithm (smooth, The MathWorks, Inc., Narick, MA, USA). This baseline was then subtracted by each individual raw spectrum. Afterwards, each spectrum is normalized to the nitrogen peak at 644.12~nm. As before, the distribution of selected peaks is analyzed by the histograms and a Wilcoxon rank sum test~(ranksum, The MathWorks, Inc., Narick, MA, USA).

\begin{figure}[ht!]
	\centering
	\includegraphics[width= \linewidth]{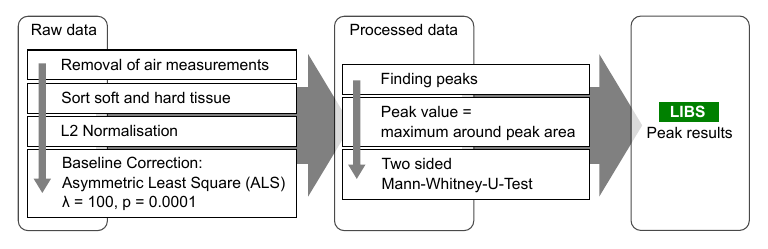}
	\caption{Schematic of the data processing algorithm used for the LIBS measurements. First, measurements that were out-of-focus are filtered out. The raw data is then normalized and sorted between soft and hard tissue based on the ratio of calcium and potassium. Afterwards, a base line correction is done, and the peak values are collected.}
	\label{processing_LIBS}
\end{figure}
The technique for processing and analysing the LIBS data is outlined in figure~\ref{processing_LIBS}. After removing artefacts and classifying into hard bone and cartilage, the spectra were normalised using L2 norm. The following step involved removing the offset via the asymmetric least square fit~(ALS)~\cite{ALS}. The chosen weight was $p_{als}$=0.0001, and a regularisation parameter of $\lambda_{als}=100$ was employed. From the processed data, the peak values were subjected to the two-sided Mann-Whitney rank test for evaluation. As a large number of peaks were tested, a significance level of 0.005 was set to avoid significant results by chance.

\subsection{Raman Spectroscopy (RS)}

Due to the promising LIBS-results, Raman measurements were added. Only samples from three patients were available. The experimental setup is presented in Figure~\ref{setup}c in a schematic manner. This setup is identical to the one we utilised in our previous study~\cite{kreiss2019diffuse} where we measured the bone and cartilage of FAI-affected tissue using DRS and Raman spectroscopy. For Raman spectroscopy, a diode laser (LASER-785-LAB-ADJ-S, Newport Corporation, Ocean Optics, Dunedin, USA) with a wavelength of 785\,\textrm{nm} was used as the excitation source. The light is coupled into a fibre of a Raman coupled fibre probe (RIP-RPB-785-SMA-SMA, Ocean Optics, Dunedin, USA) via a SMA~905 coupling connector. The back-reflected light was captured and measured with a fibre-based spectrometer (QE65000 Spectrometer, Ocean Optics, Dunedin, USA). 
\begin{figure}[ht!]
	\centering
	\includegraphics[width= \linewidth]{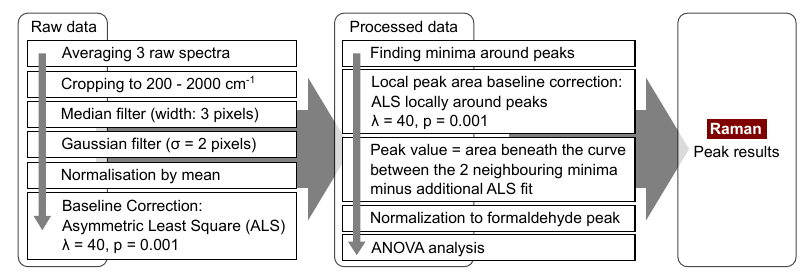}
	\caption{Schematic of the data processing algorithm used for the Raman measurements. The raw data is processed by median filtering to reduce the effect of artefacts. Afterwards, noise reduction is done by Gaussian smoothing. After a normalization step, ALS is used for the baseline subtraction of autofluorescence. Around the peaks, the minima of the spectra are found as marker for the peaks. For this area, a second ALS is done to access the real peak height. Finally, the peak area is taken, normalized to the formaldehyde peak, and an ANOVA is performed.}
	\label{fig_baseline_Raman}
\end{figure}

A total of 200 spectra were recorded, ranging from 788 to 940 nm or 52 to 2100\,\textrm{$cm^{-1}$}, with 100 obtained from the tissue affected by FAI and 100 from the healthy reference tissue.  For each spot, five spectra were acquired with an integration time of 20\,\textrm{s} each. As the Raman signal of tissue had displayed strong autofluorescence with an excitation wavelength of 785\,\textrm{nm}, a correction procedure became necessary. Figure~\ref{fig_baseline_Raman} details the implemented algorithm. Initially, the raw spectra of all three spots were averaged and then cropped to the spectral range of 200~to~2000\,\textrm{$cm^{-1}$}. A one-dimensional median filter, with a width of three pixels, was used to eliminate any artefacts from the sensor readout. The noise reduction process entailed utilising a two-pixel sigma Gaussian filter. Following this stage, the signal was normalised to the mean. The estimation of autofluorescence background was executed through ALS, with a $p_{als}$=0.001 as its asymmetric weight and a regularisation parameter of $\lambda_{als}=40$. Ultimately, the Raman signal was obtained by subtracting the autofluorescence signal.

From the processed signal, the maxima and minima were used to assign the Raman peaks. If a maximum was positioned more than 10\,\textrm{$cm^{-1}$} away, the peak was discarded. For the localized peaks, an additional ALS was conducted using a weight of $p_{als}$=0.0001 and a regularization parameter of $\lambda_{als}=0.1$ to fit a function to the minima. The value of the peak was determined by calculating the area beneath the curve between the two neighbouring minima minus additional ALS fit.

For the final analysis of Raman data, the spectra were normalized using the formaldehyde peak at 911\,\textrm{$cm^{-1}$} as the formalin peaks at 542\,\textrm{$cm^{-1}$}, 1046\,\textrm{$cm^{-1}$}, 1239\,\textrm{$cm^{-1}$}, and 1492\,\textrm{$cm^{-1}$} have considerable overlaps with other peaks. As the formaldehyde concentration was the same for all samples, these peaks are assumed to be independent of the sample under investigation. However, as the sample size varied, there may have been a slight difference in the final concentration, which can affect the Raman peaks of formaldehyde~\cite{Raman_Formalin}. The Raman peaks may not always be sufficiently clear for data analysis. Consequently, such measurements are eliminated by verifying the existence of the formaldehyde peak. By this, the total number of measurements was reduced to 170. Of these, 97 were from the FAI, and 73 from healthy tissue.

%The comparison is therefore made with Takahashi et al.~\cite{RS_effect_formalin}. 

After normalisation, the distribution of selected peaks across all spectra was analysed.  To achieve this, the statsmodel framework~\cite{seabold2010statsmodels} was used to perform an analysis of variance~(ANOVA). The three parameters considered for the analysis were FAI, sample, and position on the sample. The latter two were selected to estimate intra- and inter-patient variation. As only three patients were measured, potential significance is only assumed if the partial eta-squared of the FAI ($\partial \eta^2_{\text{FAI}}$) was at least similar to the partial eta-squared of the inter-patient variation ($\partial \eta^2_{\text{patients}}$). Furthermore, as many peaks were examined, potential significance was only accepted if p<0.0001.

\section{Conclusion}

The study demonstrates the significant potential for non-contact, all-optical and label-free analysis of tissue samples. While our previous analysis on DRS data showed the presence of pheomelanin in FAI-affected samples~\cite{kreiss2019diffuse}, our new data with RS and LIBS provide an even more detailed insight into the biochemical tissue composition, which enabled interesting discoveries on the development of FAI. The unified analysis including absorptive properties, scattering behaviour, elemental composition, and molecular binding permitted a forecast of the biochemical progression of the ailment.

It is also noteworthy that all results were obtained without markers, without direct contact with the sample and without significantly damaging it. Therefore, the suggested combination of DRS, RS and LIBS is versatile and could have significant implications for investigating of biochemical pathways in many other applications. The non-contact, all-optical and label-free approach is especially appealing for \textit{in vivo} applications, e.g., via endoscopy or during laser surgery. Additionally, all the employed technologies can operate with micrometre resolution, enabling the generation of crucial insights at a high spatial resolution.

Notwithstanding, we emphasize that this technique relies on substantial interdisciplinary knowledge. Our method can be viewed as holistic approach of three main pillars: domain knowledge of the biochemical pathways, technological understanding of optical contrast and spectroscopy, and finally, an in-depth statistical analysis.

\section*{Acknowledgement}

The authors gratefully acknowledge funding of the Erlangen Graduate School in Advanced Optical Technologies (SAOT) by the Bavarian State Ministry for Science and Art.
 
\noindent
The authors gratefully acknowledge the support of the non-profit German Arthritis Society (Deutsche Arthrose-Hilfe e.V.) and its president Helmut H. Huberti (to MD by grant P319).

\noindent
The authors would like to thank the German Research Foundation (DFG-Deutsche Forschungsgemeinschaft) for its support. This work was partly achieved in the context of the DFG-project "Kalibrierungsfreie laserinduzierte Plasmaspektroskopie (LIBS) für die Analyse der elementaren Zusammensetzung von Gewebe " (project number 502911968).

\section{Conflict of interest}
The authors declare no conflict of interest.

\section*{Data availability}

The data that support the findings of this study are available from the corresponding author upon reasonable request.

\section*{Declaration of generative AI and AI-assisted technologies}
In the writing process during the preparation of this work, the authors used "DeepL Write" in order to improve language and readability. After using this tool, the authors reviewed and edited the content as necessary and they take full responsibility for the content of the publication.

\bibliographystyle{geschichtsfrkl }
\bibliography{Reference}

\newpage
\section*{Attachment}

\subsection*{LIBS}

%Change to table

\subsubsection*{Hard tissue}

\begin{table*}[!htb]
	\centering
	%\begin{scriptsize}
	\begin{tabular}{|c|c|c|}
		\hline
		Element & Increased in FAI & p-value \\
		\hline
		C (I), 247.93 nm & False & 5.0$ \cdot 10^{-03}$ \\ 
		\hline
		Mg (II), 279.63 nm & False & 1.1$ \cdot 10^{-03}$ \\ 
		\hline
		Mg (II), 280.36 nm & False & 2.3$ \cdot 10^{-01}$ \\ 
		\hline
		Mg (I), 285.28 nm & True & 4.7$ \cdot 10^{-01}$ \\ 
		\hline
		Ca (II), 316.00 nm & True & 7.8$ \cdot 10^{-03}$ \\ 
		\hline
		Ca (II), 318.04 nm & True & 1.1$ \cdot 10^{-04}$ \\ 
		\hline
		Ca (II), 393.46 nm & True & 3.7$ \cdot 10^{-01}$ \\ 
		\hline
		Ca (II), 396.84 nm & True & 2.4$ \cdot 10^{-01}$ \\ 
		\hline
		Ca (I), 422.78 nm & True & 9.5$ \cdot 10^{-02}$ \\ 
		\hline
		Ca (I), 559.04 nm & True & 1.5$ \cdot 10^{-06}$ \\ 
		\hline
		Ca (I), 559.63 nm & True & 2.4$ \cdot 10^{-06}$ \\ 
		\hline
		Na (I), 589.15 nm & False & 5.3$ \cdot 10^{-08}$ \\ 
		\hline
		Na (I), 589.77 nm & False & 1.5$ \cdot 10^{-05}$ \\ 
		\hline
		N (I), 644.12 nm & True & 7.2$ \cdot 10^{-05}$ \\ 
		\hline
		H (I), 656.73 nm & False & 7.9$ \cdot 10^{-09}$ \\ 
		\hline
		O (I), 715.93 nm & True & 4.3$ \cdot 10^{-01}$ \\ 
		\hline
		N (I), 742.64 nm & False & 6.8$ \cdot 10^{-01}$ \\ 
		\hline
		N (I), 744.48 nm & False & 1.6$ \cdot 10^{-07}$ \\ 
		\hline
		N (I), 747.10 nm & False & 1.3$ \cdot 10^{-05}$ \\ 
		\hline
		K (I), 766.72 nm & False & 3.6$ \cdot 10^{-04}$ \\ 
		\hline
		K (I), 770.14 nm & False & 1.5$ \cdot 10^{-01}$ \\ 
		\hline
		O (I), 777.63 nm & False & 1.6$ \cdot 10^{-10}$ \\ 
		\hline
		O (I), 844.88 nm & False & 1.4$ \cdot 10^{-14}$ \\ 
		\hline
		N (I), 868.35 nm & False & 8.8$ \cdot 10^{-08}$ \\ 
		\hline
	\end{tabular}
	\caption{Detailed information about each analysed LIBS peak for the hard tissue.}
	\label{tab_LIBS_attachment_I}
	%\end {scriptsize}
\end{table*}

\subsubsection*{Soft tissue}

\begin{table*}[!htb]
	\centering
	%\begin{scriptsize}
	\begin{tabular}{|c|c|c|}
		\hline
		Element & Increased in FAI & p-value \\
		\hline
		C (I), 247.93 nm & False & 1.3$ \cdot 10^{-03}$ \\ 
		\hline
		Mg (II), 279.63 nm & True & 8.2$ \cdot 10^{-01}$ \\ 
		\hline
		Mg (II), 280.36 nm & False & 7.3$ \cdot 10^{-02}$ \\ 
		\hline
		Mg (I), 285.28 nm & False & 9.6$ \cdot 10^{-01}$ \\ 
		\hline
		Ca (II), 316.00 nm & False & 9.9$ \cdot 10^{-01}$ \\ 
		\hline
		Ca (II), 318.04 nm & False & 9.5$ \cdot 10^{-01}$ \\ 
		\hline
		Ca (II), 393.46 nm & False & 2.3$ \cdot 10^{-04}$ \\ 
		\hline
		Ca (II), 396.84 nm & False & 1.9$ \cdot 10^{-04}$ \\ 
		\hline
		Ca (I), 422.78 nm & False & 3.0$ \cdot 10^{-03}$ \\ 
		\hline
		Ca (I), 559.04 nm & False & 6.9$ \cdot 10^{-01}$ \\ 
		\hline
		Ca (I), 559.63 nm & False & 6.5$ \cdot 10^{-01}$ \\ 
		\hline
		Na (I), 589.15 nm & False & 9.7$ \cdot 10^{-01}$ \\ 
		\hline
		Na (I), 589.77 nm & True & 4.8$ \cdot 10^{-01}$ \\ 
		\hline
		N (I), 644.12 nm & False & 6.4$ \cdot 10^{-01}$ \\ 
		\hline
		H (I), 656.73 nm & False & 3.5$ \cdot 10^{-02}$ \\ 
		\hline
		O (I), 715.93 nm & False & 8.8$ \cdot 10^{-01}$ \\ 
		\hline
		N (I), 742.64 nm & True & 3.5$ \cdot 10^{-02}$ \\ 
		\hline
		N (I), 744.48 nm & False & 8.2$ \cdot 10^{-01}$ \\ 
		\hline
		N (I), 747.10 nm & True & 4.7$ \cdot 10^{-01}$ \\ 
		\hline
		K (I), 766.72 nm & True & 1.5$ \cdot 10^{-02}$ \\ 
		\hline
		K (I), 770.14 nm & True & 7.3$ \cdot 10^{-02}$ \\ 
		\hline
		O (I), 777.63 nm & False & 1.6$ \cdot 10^{-01}$ \\ 
		\hline
		O (I), 844.88 nm & True & 7.5$ \cdot 10^{-01}$ \\ 
		\hline
		N (I), 868.35 nm & False & 3.3$ \cdot 10^{-01}$ \\ 
		\hline
	\end{tabular}
	\caption{Detailed information about each analysed LIBS peak for the soft tissue. }
	\label{tab_LIBS_attachment_II}
	%\end {scriptsize}
\end{table*}

\newpage

\subsection*{Raman}

\begin{table*}[!htb]
	\centering
	%\begin{scriptsize}
		\begin{tabular}{|c|c|c|c|c|c|c|}
				\hline
				Component & df$_{FAI}$ & Sum sq. (FAI)& Mean sq (FAI) & F$_{FAI}$ & p$_{FAI}$ & $\partial \eta ^2 _{FAI}$ \\
				\hline
				\textbf{Hydroxyproline} & 2 & 27 & 14 & 34 & 1.5$ \cdot 10^{-13}$ & 0.25 \\
				\hline
				C-C stretch & 2 & 8.1 & 4.1 & 20 & 9.8$ \cdot 10^{-9}$ & 0.17 \\
				\hline
				\textbf{$\bm{v_3 PO_4 ^{3}}$} & 2 & 230 & 120 & 190 & 2.1$ \cdot 10^{-47}$ & 0.65  \\
				\hline
				\textbf{Phenylalanine} & 2 & 2500 & 1300 & 130 & 1.1$ \cdot 10^{-36}$ & 0.56 \\
				\hline
				\textbf{Pyranose ring} & 2 & 150 & 76 & 130 & 4.1$ \cdot 10^{-36}$ & 0.55 \\
				\hline
				$\nu_1 CO_3^{2-}$ & 2 & 39 & 20 & 17 & 1.6$ \cdot 10^{-7}$ & 0.14 \\
				\hline
				\textbf{Pyranose ring} & 2 & 43 & 22 & 120 & 9.5$ \cdot 10^{-35}$ & 0.54 \\
				\hline
				\textbf{Amide III} & 2 & 79 & 39 & 51 & 1.5$ \cdot 10^{-18}$ & 0.33 \\
				\hline
				Lipid $CH_2$ twist & 2 & 1.6 & 0.78 & 7.4 & 0.00079 & 0.07 \\
				\hline
				Tyrosine & 2 & 0.24 & 0.12 & 1.9 & 0.15 & 0.02 \\
				\hline
				\textbf{Amide III GAGs} & 2 & 39 & 19 & 110 & 5.1$ \cdot 10^{-33}$ & 0.52  \\
				\hline
				\textbf{$\bm{CH_2}$ bend} & 2 & 96 & 48 & 108 & 8.1$ \cdot 10^{-33}$ & 0.52 \\
				\hline
				Phenylring  & 2 & 0.49 & 0.25 & 5.4 & 0.0052 & 0.05 \\
				\hline
				Tyrosine & 2 & 40 & 3.0 & 27 & 3.4$ \cdot 10^{-11}$ & 0.04 \\
				\hline
				Amide I & 2 & 1.8 & 0.88 & 6.5 & 0.0019 & 0.06 \\
				\hline
				\textbf{Carbonyl} &  2 & 12 & 5.8 & 81 &1.0$ \cdot 10^{-26}$ & 0.45 \\
				\hline
		\end{tabular}
	\caption{Information about the ANOVA for the FAI parameter. The bold components are seen as statistical relevant.}
	\label{tab_Raman_attachment}
	%\end {scriptsize}
\end{table*}

\end{document}